\documentclass[conference]{IEEEtran}
\IEEEoverridecommandlockouts
 \usepackage[table,xcdraw]{xcolor}
\usepackage{amsmath}
\usepackage{amsfonts}
\usepackage{framed}
\usepackage{graphicx}
\usepackage{textcomp}
\usepackage{xcolor}
\usepackage{multirow}
\usepackage{soul}
\usepackage{booktabs}
\usepackage{tikz}
\usepackage{algorithm, enumitem}
\usepackage{algpseudocode}
\usepackage{colortbl}
\usepackage{bm}
\usepackage{url}
\usepackage{float}
\usepackage{hyperref}
\usepackage{filecontents}
\usepackage{mdframed}
\usepackage{todonotes}
\newcommand{\cmmnt}[1]{}  

\usepackage{setspace}
\usepackage{pifont}% http://ctan.org/pkg/pifont
\makeatletter 
\newcommand\semiHuge{\@setfontsize\semiHuge{22.72}{27.38}}
\makeatother

\usepackage[a4paper, total={184mm,239mm}]{geometry}
\def\BibTeX{{\rm B\kern-.05em{\sc i\kern-.025em b}\kern-.08em
    T\kern-.1667em\lower.7ex\hbox{E}\kern-.125emX}}
\begin{document}

\title{\semiHuge CoViS: Compact Token Streaming from Sensor to Attention for Energy-Efficient Vision\vspace{-2mm}}
\author{\IEEEauthorblockN{Chengwei Zhou, Khan Md Zobayer Hassan, Hassan Najafi, Hossein Lavasani, and Gourav Datta}\\
\IEEEauthorblockA{Case Western Reserve University, Cleveland, Ohio, USA}}

\maketitle

\begin{abstract}
High-resolution, high-frame-rate vision sensors generate data volumes whose
transfer and processing impose substantial energy overhead. Near-sensor
computing extracts compact features at the sensor, but aggressive extraction
raises sensor-side compute, memory, and power, and reconstructing compressed
features for conventional backbones erodes end-to-end savings. We propose
CoViS, a sensor-to-processor vision architecture that jointly optimizes
near-sensor computation, communication, and inference. Multi-scale
frequency-domain tokenization with depthwise-separable processing cuts
sensor-output bandwidth up to $85.3\times$ using $3.1$\,M sensor-side MACs;
transmitted features are expanded once and consumed directly by a lightweight
attention backend, holding processor computation to $29.0$\,M MACs. For
always-on video, fixed-rate spatial sampling transmits a deterministic token
subset per frame while preserving full spatial coverage over time. CoViS
lowers system energy up to $281.9\times$ over raw-frame inference on BLE at competitive accuracy, showing that representation, communication, and
inference must be co-designed rather than optimizing bandwidth in isolation.
\end{abstract}

\begin{IEEEkeywords}
Near-sensor, energy-efficient, hardware--algo-rithm
co-design, visual feature compression, always-on sensing.
\vspace{-5mm}
\end{IEEEkeywords}

\section{Introduction}
\vspace{-1mm}
Modern visual intelligence increasingly relies on high-resolution and high-frame-rate cameras for applications ranging from smart-home monitoring to autonomous and embedded systems. In conventional architectures, captured images or video frames are transferred from the sensor to a separate processor for neural-network inference. As resolution and frame rate increase, this continuous movement of high-dimensional visual data incurs substantial communication bandwidth and energy, motivating in-/near-sensor computing, where early processing is moved closer to image acquisition and only an intermediate representation is transmitted downstream \cite{zhou2020nearsensor,chai2023insensor}.

A growing body of work has explored this paradigm using shallow neural networks, processing-in-pixel architectures, mixed-signal computation, and near-sensor feature extraction \cite{xu2021senputing,bose2020scamp,abedin2022mrpipa,xu2020dpce,datta2022p2m,mennel2020ultrafast}. These works demonstrate that moving computation toward the sensor can substantially reduce the amount of data leaving the focal plane. However, they also expose an important architectural tradeoff. 
% Performing more feature extraction near the sensor generally enables stronger representation compression, but increases compute, memory, area, and power requirements within the tightly constrained logic integrated with the image sensor. Lightweight processing reduces this implementation burden, but may provide insufficient data reduction to justify the added complexity of near-sensor integration. 
More aggressive near-sensor feature extraction enables stronger compression but increases compute, memory, area, and power requirements, whereas lightweight processing may yield insufficient data reduction.
%The second challenge appears after the representation leaves the sensor. 
Many prior approaches optimize the amount of data crossing the sensor-processor interface while retaining compatibility with conventional downstream networks \cite{abedin2022mrpipa,datta2022p2m,zhou2020nearsensor}. This often requires the compressed representation to be projected or expanded before entering the original backbone \cite{zhou2026frequencyformer}. %For example, FrequencyFormer work \cite{zhou2026frequencyformer} performs aggressive frequency-domain compression near the sensor, but reconstructs the compact representation to the token dimensions expected by a standard ViT backbone. 
While this preserves compatibility with pretrained models, much of the downstream computation remains unchanged.

\begin{figure}
    \centering
    \includegraphics[width=\linewidth]{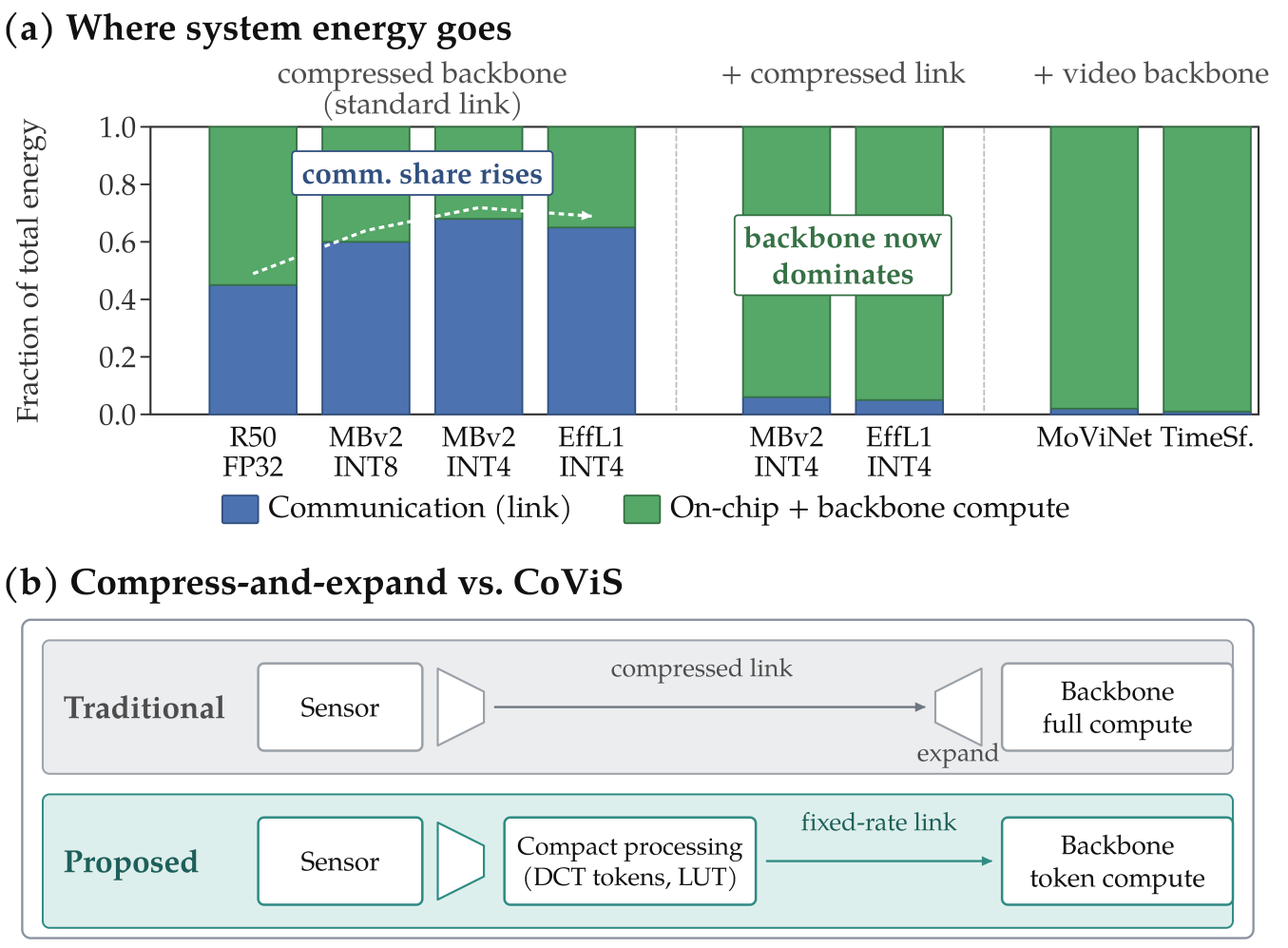}
    \vspace{-9mm}
\caption{\footnotesize \textbf{(a)} Compressing the backbone raises the communication share;
compressing the link moves the bottleneck to the backbone. \textbf{(b)} Conventional pipelines expand before a full-width backbone;
CoViS stays compact through link and backbone.}
    \label{fig:motiv}
    \vspace{-5mm}
\end{figure}
These challenges are especially considered in wireless vision systems, such as wearable cameras and distributed visual sensors, where creates an Amdahl's-law bottleneck at the system level. Profiling CNN,
transformer, and video backbones over a wireless interface
(Fig.~\ref{fig:motiv}a) shows communication rising to $45$--$68\%$ of system
energy as the backbone is compressed; an efficient link then leaves backend inference dominant. Reducing communication further yields diminishing end-to-end benefit unless the compact representation also cuts downstream computation and activation movement. The gap widens for streaming video, where most system energy is spent after sensor-side feature extraction. 
We therefore ask a broader architectural question: \textit{can a representation be inexpensive to generate near the sensor, compact to transmit, and efficient to directly compute with, while preserving sufficient information for downstream inference?}

To address this problem, we propose CoViS, an end-to-end sensor-to-processor vision architecture that keeps visual representations compact across sensing, communication, and inference, rather than compressing features solely for transmission. The sensor performs multi-scale DCT tokenization, lightweight depthwise-separable stems, and FiLM, producing compact token maps. 
Their spatial resolution, feature width, and precision jointly determine the sensor workload, communication payload, and backend input, rather than being optimized for compression alone.
These features are transmitted directly and projected once to a narrow width
the attention backend is built around, rather than restored to a pretrained
backbone's dimensions (Fig.~\ref{fig:motiv}b).
For always-on video, CoViS introduces fixed-rate spatial sampling, which transmits a subset of the token grid per frame, maintaining payload while preserving full spatial coverage over time. 
Our contributions are:
\vspace{-1mm}
\begin{itemize}[leftmargin=*]
\item \textbf{Lightweight near-sensor encoding:}
We propose multi-scale DCT tokenization with depthwise-separable stems to produce compact visual tokens within sensor compute constraints. This reduces sensor-output bandwidth by up to $85.3\times$ with only $3.1$\,M sensor-side MACs.

% \item \textbf{Fixed-rate streaming:}
% We introduce a deterministic spatial sampling schedule that transmits a fixed token budget per frame while covering the full field of view over time, enabling space--time attention at a constant communication rate.
% \item \textbf{Direct inference and Representation--communication--inference co-design:}
% We jointly design the transmitted representation and a compact attention backend to reduce both communication and downstream computation. CoViS requires only $29.0$\,M processor MACs and reduces system energy by up to $281.9\times$ via BLE link, relative to raw-frame inference.

\item \textbf{Direct inference and system-level co-design:}
We propose a lightweight attention backend that consumes the compact tokens without reconstructing a full-width feature map. Across the sensor, link, and processor, CoViS requires $29.0$\,M processor MACs and reduces system energy by up to $281.9\times$ versus raw-frame inference over BLE.

\item \textbf{Fixed-rate streaming for always-on vision:}
We introduce deterministic periodic spatial sampling with
factorized space--time attention, jointly reducing communication and video inference costs while ensuring predictable communication loads and full spatial coverage over time.

\end{itemize}

\begin{figure}
    \centering
    \includegraphics[trim={160 179 330 157}, clip, width=\linewidth]{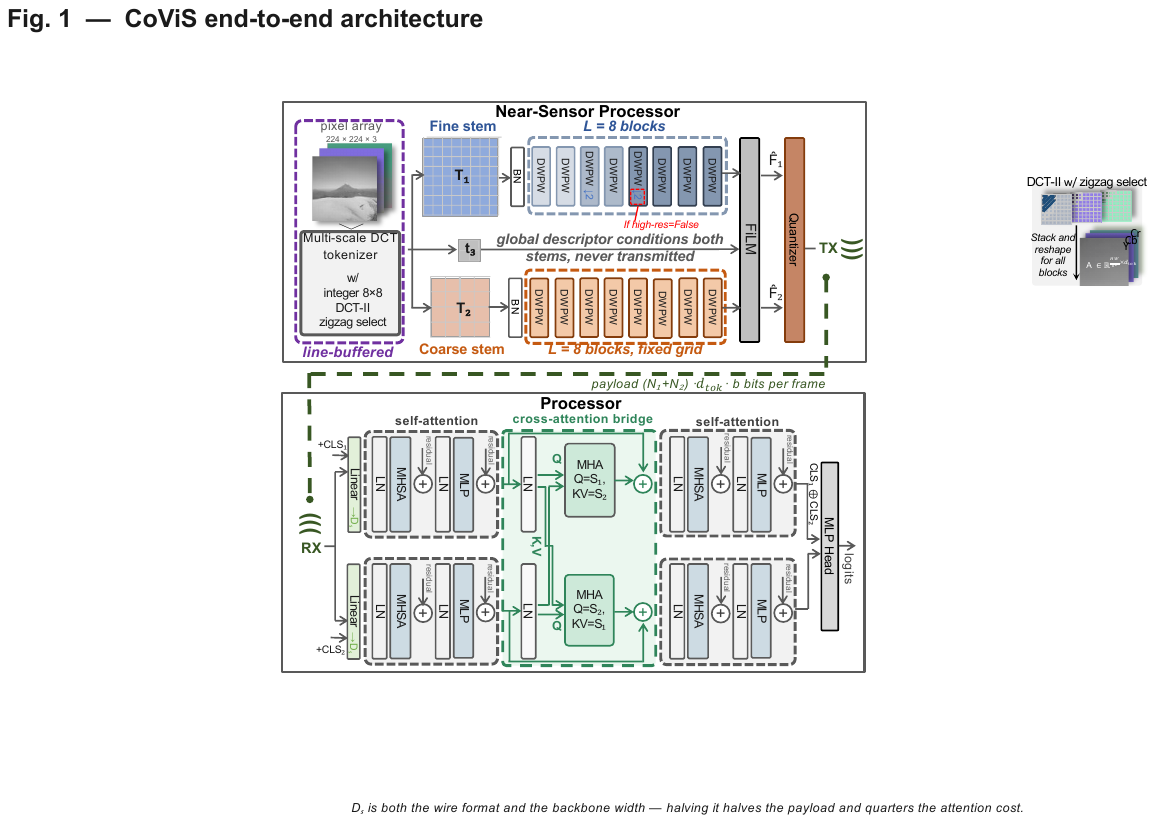}
    \vspace{-7mm}
    \caption{CoViS end-to-end architecture. The sensor node runs the whole
    feature-extraction path and transmits two quantized token maps of width
    $d_{\mathrm{tok}}$ over a low-power wireless link; the processor projects once to $D_s$
    and then only attends. The global descriptor $\mathbf{t}_3$ is consumed by FiLM on the
    sensor and never leaves it.}
    \label{fig:arch}
    \vspace{-4mm}
\end{figure}

\vspace{-1mm}
\section{Related Work}
\vspace{-1mm}
\noindent\textbf{Frequency-Domain Feature Encoding}: Frequency transforms such as DCT provide compact visual representations through energy compaction~\cite{wallace1992jpeg}, motivating their use in CNNs~\cite{chen2016packing,gueguen2018faster,ehrlich2019deep,xu2020learning} and transformers through spectral mixing, frequency-aware attention, and DCT-based tokenization~\cite{lee2022fnet,qin2021fcanet,patro2025spectformer,li2023dcformer,pan2024dctattention,lee2024dctvit}. Other approaches reduce representation size through learned compression~\cite{balle2017end,balle2018variational} or token pruning and merging~\cite{rao2021dynamicvit,bolya2023token}. Closest to our setting, Zhou et al.~\cite{zhou2026frequencyformer} performs DCT-based compression at the sensor but restores the feature width expected by a pretrained ViT, leaving the backbone unchanged. In contrast, CoViS jointly reduces the transmitted width and processor computation, matching the representation to the tight compute, area, and bandwidth budgets of the sensor node.

\noindent\textbf{Energy-Efficient Data Transfer}: Recent systems co-design sensing, compression, and communication to reduce camera data movement~\cite{GT_Comp,photonic_in-sensor,ISSCC21_Comm,JSSC14_Comm}. While sensor-side processing can achieve $1$--$20~\mathrm{pJ/pixel}$, subsequent data transfer may still exceed $1~\mathrm{nJ/pixel}$. Measured BLE-class radios further illustrate this gap: a $0$\,dBm transmitter consumes $4.5$\,mW~\cite{sharifzadeh2018fully}, while a direct-conversion receiver consumes $980$\,\textmu W~\cite{shirazi2021ultralow}. Existing designs mainly reduce the transmitted volume while leaving the receiving network unchanged, with the resulting rate varying with content. 
We instead jointly design the transmitted feature width and backbone architecture, while adopting deterministic periodic sampling to ensure predictable communication loads for always-on operation.

\vspace{-1mm}
\section{Method}
\label{sec:method}
\vspace{-1mm}

The transmitted representation should be inexpensive to extract, compact for communication, and informative enough for downstream inference. Existing near-sensor pipelines often compress features before transmission but expand them again to match a pretrained backbone, limiting downstream savings. CoViS (Fig~\ref{fig:arch}) instead transmits compact features at reduced size, expands them once on the processor, and applies attention directly. Trained end-to-end from scratch, the entire pipeline jointly optimizes sensor-side compute, communication, and backend inference. The following subsections detail the near-sensor encoder, interface, and processor backend.

\vspace{-1mm}
\subsection{Near-Sensor Feature Extraction}
\label{sec:stems}
\vspace{-1mm}

\subsubsection{Multi-scale DCT tokenization}
\label{sec:tokenizer}

\begin{figure}
    \centering
    \includegraphics[trim={160 225 355 285}, clip, width=1\linewidth]{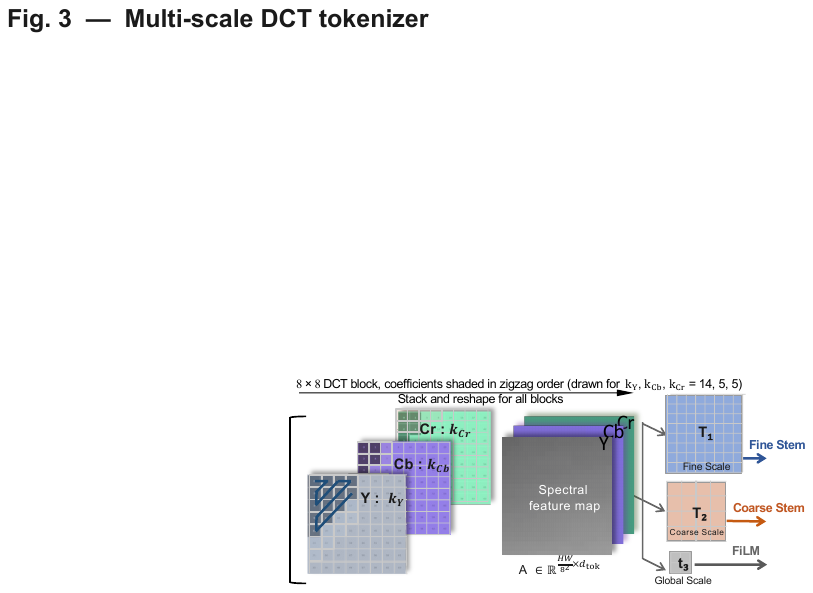}
    \vspace{-8mm}
    \caption{Multi-scale DCT tokenizer. Zigzag selection keeps the first $k_Y$, $k_{Cb}$ and
    $k_{Cr}$ coefficients of each block, giving the feature map $\mathbf{A}$ of width $d_{\mathrm{tok}}$. Three scales are read from it and feed the
    fine stem, the coarse stem and FiLM.}
    \label{fig:tokenizer}
    \vspace{-6mm}
\end{figure}
 
The input is first converted to YCbCr color space, following standard practice~\cite{wallace1992jpeg,ibraheem2012ycbcr}. Then the Y, Cb, and Cr channels are each partitioned into non-overlapping $8{\times}8$ blocks, and the 2D DCT is applied independently to each block. This block size is chosen for compatibility with the JPEG standard and captures fine-grained local texture and edge information. For an $N{\times}N$ spatial block $\mathbf{X}$, the 2-D DCT computes the coefficient matrix
$
  \mathbf{Y} = \mathbf{C}\,\mathbf{X}\,\mathbf{C}^{\!\top},
  \label{eq:dct2d}
$
where $\mathbf{C}\!\in\!\mathbb{R}^{N\times N}$ is the orthonormal DCT basis with entries
$
  C_{k,i} = w_k\,\cos\!\Bigl(\tfrac{\pi(2i{+}1)k}{2N}\Bigr),
  \quad
  w_k =
  \begin{cases}
    \sqrt{1/N} & k=0,\\
    \sqrt{2/N} & k>0.
  \end{cases}
  \label{eq:dct_basis}
  \vspace{-1mm}
$
For a $H{\times}W$ single-channel image, $8{\times}8$ block-wise DCT produces a grid of $\frac{H}{8}{\times}\frac{W}{8}$ blocks, each containing $64$ frequency coefficients. After channel selection, we retain $k_Y$, $k_{Cb}$, and $k_{Cr}$ coefficients from Y, Cb, and Cr respectively, yielding a feature map $\mathbf{A}$ of size $\frac{H}{8}{\times}\frac{W}{8}{\times}(d_{\mathrm{tok}} = k_Y + k_{Cb} + k_{Cr})$. We make $(k_Y,k_{Cb},k_{Cr})$ configurable rather than fixed, since these parameters jointly determine how much detail is preserved and the required stem width. The tokenizer then processes the output through three parallel scales to capture multi-scaled spatial-frequency features.
 
\noindent\textbf{Channel selection.} Natural image blocks concentrate most of their energy in low-frequency coefficients, allowing higher-frequency channels to be discarded with limited information loss. The zigzag scan visits the $64$ entries of $\mathbf{Y}$ from low to high frequency, and we keep only its first few steps (see Fig.~\ref{fig:tokenizer}). Writing $\mathbf{Y}^{Y}_{p,q}$, $\mathbf{Y}^{Cb}_{p,q}$, and $\mathbf{Y}^{Cr}_{p,q}$ for the coefficient matrices of block $(p,q)$ in Y, Cb, and Cr, we have
$
\mathbf{A}[p,q,:] = \big[\,\zeta_{k_Y}(\mathbf{Y}^{Y}_{p,q}),\;
\zeta_{k_{Cb}}(\mathbf{Y}^{Cb}_{p,q}),\;
\zeta_{k_{Cr}}(\mathbf{Y}^{Cr}_{p,q})\,\big],
\label{eq:bandsel}
$
where $\zeta_k(\cdot)$ returns the first $k$ coefficients of the scan. We keep more of Y than of Cb and Cr, because the eye resolves color less sharply than brightness. This brings a block position down from $192$ coefficients to $d_{\mathrm{tok}}$, before any learned layer sees it.
 
To capture frequency information at multiple spatial extents without introducing additional transforms, we construct three complementary scales from the same block-DCT representation: a fine scale for local structure, a coarse scale for regional context, and a global scale for image-level spectral statistics.

\noindent\textbf{Fine scale.} The fine-scale representation directly uses $\mathbf{A}$:
\vspace{-1mm}
\begin{equation}
  \mathbf{T}_1 = \mathbf{A}
  \in \mathbb{R}^{\frac{HW}{8^2}\times d_{\mathrm{tok}}},
  \label{eq:t1}
  \vspace{-2mm}
\end{equation}
thereby preserving the native $8{\times}8$ block grid and its local frequency information.

\noindent\textbf{Coarse scale.} A larger spatial support captures structures that may not be represented within a single $8{\times}8$ block. Rather than computing a separate $32{\times}32$ DCT, we average the $8{\times}8$ coefficients over groups of $4{\times}4$ neighboring blocks, yielding
\vspace{-1mm}
\begin{equation}
  \mathbf{T}_2 = \mathrm{AvgPool}_4(\mathbf{A})
  \in \mathbb{R}^{\frac{HW}{32^{2}}\times d_{\mathrm{tok}}},
  \label{eq:t2}
  \vspace{-2mm}
\end{equation}
which provides a coarse-scale representation with substantially lower computation than an additional $32{\times}32$ block transform.

\noindent\textbf{Global scale.} At the image level, we average $\mathbf{A}$ across all spatial blocks:
\vspace{-2mm}
\begin{equation}
  \mathbf{t}_3 = \frac{64}{HW}\sum_{p,q}\mathbf{A}[p,q,:]
  \;\in\;\mathbb{R}^{d_{\mathrm{tok}}}.
  \label{eq:t3}
  \vspace{-2mm}
\end{equation}
This produces one value per retained frequency channel and summarizes the image-level distribution of spectral energy.

None of these scale-construction steps introduces learnable parameters, so the tokenizer requires no model-weight storage on the sensor. With an integer approximation of the DCT, the complete tokenizer can also be implemented using integer arithmetic. Since \eqref{eq:bandsel} retains only a subset of frequency coefficients, only the corresponding rows of $\mathbf{C}$ need to be evaluated. The implementation requires a single eight-row line buffer, allowing tokenization to proceed during pixel readout without buffering the complete frame. The three scales are then used differently by the sensor-side network. After batch normalization, $\mathbf{T}_1$ is processed by the high-resolution stem and $\mathbf{T}_2$ by the low-resolution stem, while $\mathbf{t}_3$ is provided to a FiLM layer that conditions both stems.

\subsubsection{Two depthwise-separable stems \& FiLM}
 
Both stems are $L$ stacks of the same residual block,
\vspace{-1mm}
\begin{equation}
  \mathrm{DWPW}(\mathbf{Z}) = \mathbf{Z} +
  \mathrm{PW}_{1\times1}\!\big(\sigma(\mathrm{BN}(\mathrm{DW}_{3\times3}(\mathbf{Z})))\big),
  \label{eq:dwpw}
  \vspace{-1mm}
\end{equation}
where $\mathbf{Z}$ is the hidden state, $\mathrm{DW}_{3\times3}$ filters each channel on its own, $\mathrm{PW}_{1\times1}$
mixes channels, and $\sigma$ is GELU. When the width changes we add a $1{\times}1$
convolution on the skip path.
 
Two things keep these stems small enough for the sensor. The considerable spatial reduction of the pixel array ($H \times W$), together with \eqref{eq:bandsel}, has already reduced the channel count to $d_{\mathrm{tok}}$, so each layer operates on a small tensor. Splitting a $3{\times}3$ convolution into a per-channel filter followed by a $1{\times}1$ channel-mixing layer further reduces the computation required for both spatial filtering and channel mixing.

\noindent\textbf{Fine Stem}. We interpret $\mathbf{T}_1$ as a $\frac{H}{8}\times\frac{W}{8}\times d_{\mathrm{tok}}$ feature map. The Fine Stem first applies two $\mathrm{DWPW}$ blocks at this resolution, followed by a stride-2 $2{\times}2$ convolution that halves the spatial resolution. For the high-resolution option (spatial reduction scale $s=16$), we then apply six additional $\mathrm{DWPW}$ blocks at the resulting resolution. For the most aggressive $s=32$ option, we instead apply two $\mathrm{DWPW}$ blocks, followed by another stride-2 $2{\times}2$ convolution and four additional $\mathrm{DWPW}$ blocks.

The output $\mathbf{F}_1$ therefore contains
$
N_1 =
\begin{cases}
14 \times 14 & s = 16\ \text{(high-res)}, \\
7 \times 7 & s = 32,
\end{cases}
$
spatial positions, considering a frame input $H$$\times$$W$=$224$$\times$$224$, each represented by a $d_{\mathrm{tok}}$-dimensional feature.

\noindent\textbf{Coarse stem}.
The coarse stem takes $\mathbf{T}_2$ and applies $L$ blocks on a fixed $\frac{H}{32}\times\frac{W}{32}$ grid at width $d_{tok}$. Its output $\mathbf{F}_2$ carries $N_2 = 7\times 7 = 49$ positions and supplies the scene-level context that the fine stem, confined to a local view, cannot reach.
 
\noindent\textbf{FiLM conditioning.}
Batch normalization in Eq. \eqref{eq:dwpw} normalizes channel-wise statistics and can reduce scene-level information encoded in $\mathbf{t}_3$. We reintroduce this global context through a FiLM layer,
\vspace{-1mm}
\begin{equation}
  [\bm{\gamma},\bm{\beta}] = \mathbf{W}_f\,\mathbf{t}_3,\qquad
  \hat{\mathbf{F}} = \mathbf{Z}^{final}\odot(1+\bm{\gamma}) + \bm{\beta},
  \label{eq:film}
  \vspace{-2mm}
\end{equation}
which is applied to the final outputs of both stems, $\mathbf{Z}_1^{final}$ and $\mathbf{Z}_2^{final}$. We initialize
$\mathbf{W}_f\in\mathbb{R}^{2d_{\mathrm{tok}}\times d_{\mathrm{tok}}}$ to zero, so the FiLM layer initially behaves as an identity mapping and subsequently learns a residual, scene-dependent modulation. Its overhead is limited to a single small matrix--vector multiplication per frame. Applying FiLM on the near-sensor side also avoids transmitting $\mathbf{t}_3$ separately. Since the global representation is used only for conditioning, Eq. \eqref{eq:film} incorporates its information directly into the two stem outputs, leaving only $\hat{\mathbf{F}}_1$ and $\hat{\mathbf{F}}_2$ to be transmitted.

 \vspace{-1mm}
\subsection{Interface}
\label{sec:interface}
\vspace{-1mm}
 
What leaves the sensor is the pair $(\hat{\mathbf{F}}_1,\hat{\mathbf{F}}_2)$. Each tensor is quantized with a learned uniform quantizer,
$
\hat{z}=s\cdot\mathrm{clip}\!\left(\left\lfloor z/s \right\rceil,
-2^{b-1},\,2^{b-1}-1\right),
$
All sensor-side operations are executed in INT8, and only the final outputs are reduced to $b$ bits for transmission; the processor then operates in higher precision of FP16. For frame rate $f$, the payload and link rate are
\vspace{-1mm}
\begin{equation}
  B_{\mathrm{frame}}=(N_1+N_2)d_{\mathrm{tok}}b,
  \qquad
  R=fB_{\mathrm{frame}}.
  \label{eq:rate}
\vspace{-1mm}
\end{equation}
Thus, communication cost is jointly controlled by the spatial token count $N_1$, feature width $d_{\mathrm{tok}}$, and precision $b$.

\vspace{-1mm} 
\subsection{Compact-Domain Backend}
\label{sec:backend}
\vspace{-1mm}
 
The processor dequantizes the two tensors and attends over them directly after expanding their feature dimension to $D_s$ to provide sufficient model capacity. There is no stem or additional spatial projection, and the feature width remains unchanged thereafter. As a result, the processor still benefits from reduced FLOPs and energy consumption.

\subsubsection{Two sequences and a bridge}
 
We give each scale its own class token and position embedding after a Linear projection to $D_s$: 
$
\vspace{-1mm}
  \mathbf{S}_1\in\mathbb{R}^{(N_1+1)\times D_s},
  \mathbf{S}_2\in\mathbb{R}^{(N_2+1)\times D_s},
$
and run pre-norm transformer blocks on each. We keep the two sequences apart rather than concatenating them, for two reasons. Attention grows with the square of sequence length, and $N_1^2+N_2^2$ is smaller than $(N_1{+}N_2)^2$. The two scales also have different spatial statistics, so a shared position embedding would have to describe both. The sequences exchange information once, through a bridge in both directions:
\begin{align}
\vspace{-1mm}
\mathbf{S}_1 &\leftarrow \mathbf{S}_1 +
  \mathrm{MHA}\big(\mathrm{LN}(\mathbf{S}_1),\,\mathrm{LN}(\mathbf{S}_2)\big),\\
\mathbf{S}_2 &\leftarrow \mathbf{S}_2 +
  \mathrm{MHA}\big(\mathrm{LN}(\mathbf{S}_2),\,\mathrm{LN}(\mathbf{S}_1)\big),
\vspace{-2mm}
\end{align}
where $\mathrm{MHA}(Q,KV)$ draws queries from its first argument. With $A$ self-attention layers per sequence, $\lceil A/2\rceil$ run before the bridge and the rest after it. Each sequence thus builds its own representation first, then shares it, then uses what it received. The classifier reads both class tokens as
$
  \mathbf{y} = \mathrm{MLP}\big([\,\mathrm{LN}(\mathbf{S}_1)_{\mathrm{cls}}
  \,\|\, \mathrm{LN}(\mathbf{S}_2)_{\mathrm{cls}}\,]\big).
$
 
% \subsubsection{Why the cut point matters}
 
% Attention costs $O(N D_s^2)$ and the link costs $O(N D_s b)$. The same $D_s$ appears in
% both, because we cut the pipeline where the features reach that width. Halving $D_s$
% therefore halves the radio payload and divides attention cost by four, and nothing in
% between undoes either saving. A design that sends a narrow code and then widens it on
% the processor gets the first saving and loses the second. Section~\ref{sec:results} measures
% this directly, by adding a projection layer at the processor and changing nothing else.

 \vspace{-1mm}
\subsection{Streaming Recognition}
\label{sec:stream}
\vspace{-1mm}

\begin{figure}
    \centering
    \includegraphics[trim={285 335 220 150}, clip, width=\linewidth]{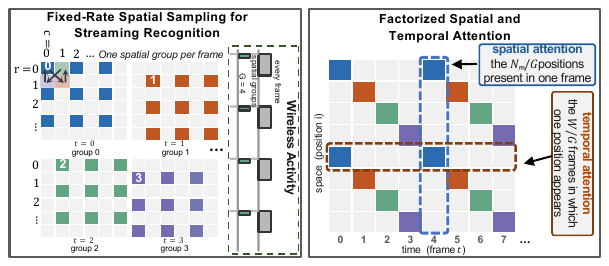}
    \vspace{-7mm}
    \caption{Fixed-rate spatial sampling for streaming recognition.
    \textbf{Left:} one of four interleaved spatial groups is transmitted per frame,
    providing full-field coverage at a deterministic periodic sampling payload.
    \textbf{Right:} tokens collected over $W$ frames are processed with factorized
    spatial and temporal attention.}
    \label{fig:placeholder}
    \vspace{-5mm}
\end{figure}

While image inference transmits a fixed representation per frame, video contains substantial temporal redundancy that can be exploited to further reduce communication. Rather than using content-dependent token selection, which produces a variable payload, we transmit a fixed fraction of spatial tokens in each frame using a deterministic schedule.

\noindent\textbf{Fixed-rate spatial sampling.}
We partition the positions of each feature stream into $G$ interleaved groups and transmit one group per frame in cyclic order. For $G{=}4$, 
$
  j(r,c) = \mathbf{B}\big[r \bmod 2,\, c \bmod 2\big],
  \mathbf{B} = \begin{bmatrix} 0 & 2 \\ 3 & 1 \end{bmatrix},
$
Each group spans the full field of view at one-quarter spatial density, while all positions are covered over four frames. The ordering further makes each pair $(0,1)$ and $(2,3)$ form a half-density quincunx pattern. Since exactly one group is transmitted per frame,
$
B_{\mathrm{frame}} = \frac{(N_1+N_2)d_{\mathrm{tok}}b}{G}, 
R = fB_{\mathrm{frame}}.
$

\noindent\textbf{Factorized attention.}
For video, the processor collects the tokens delivered over a window of $W$ frames and infers once per window, so a decision rests on $W/f$ seconds of evidence. Therefore, we exploit that split rather than attending over $W$ phases jointly. Each block applies self-attention twice, once within a frame over the $N_m/G$ ($m \in \{1,2\}$) positions present in it and once within a position over its $W/G$ occurrences, followed by the MLP. Joint attention would cost $O\!\big((WN_m/G)^2 D_s\big)$ pairwise terms, which grows with the product of window length and grid size; the factorization replaces this with
$
  O\!\Big(W\big(\tfrac{N_m}{G}\big)^{2} D_s
   \;+\; N_m\big(\tfrac{W}{G}\big)^{2} D_s\Big),
$
which decouples spatial and temporal interactions, avoiding the quadratic dependence on the product of spatial and temporal token counts. 

The two streams are fused by the same bridge as before, applied frame by frame: both streams deliver tokens in every frame, so the cross-attention at frame $t$ runs between the $N_1/G$ and $N_2/G$ positions present in it. The classifier reads the mean of each stream's window tokens in place of a class token.

\begin{figure}
     \centering
     \includegraphics[width=0.9\linewidth]{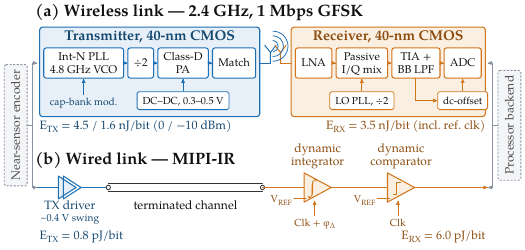}
     \vspace{-4mm}
     \caption{\footnotesize Sensor--processor interfaces. \textbf{(a)} Measured $40$-nm BLE-class
wireless link~\cite{sharifzadeh2018fully,shirazi2021ultralow} and \textbf{(b)}
wired MIPI-IR link~\cite{zhou2026frequencyformer,Int_Rx}. Both carry the same
quantized tensors; per-bit energy differs by roughly three orders of
magnitude.}
     \label{fig:txrx}
     \vspace{-5mm}
\end{figure}

 \vspace{-1mm}
\subsection{TX \& RX Communication}
\label{sec:txrx}
\vspace{-1mm}

The compact representation produced by CoViS can be transported through either a wired or wireless sensor--processor interface, allowing the same architecture to support different deployment scenarios. For tightly coupled camera--processor systems, we adopt the low-power asymmetric MIPI interface of Zhou et al.~\cite{zhou2026frequencyformer}, which shifts circuit complexity from the sensor-side transmitter to an integrating receiver, enabling reliable communication at reduced Tx swing and power. For wearable and distributed always-on systems, we instead consider a low-power $2.4$\,GHz wireless link based on measured BLE-class transmitter~\cite{sharifzadeh2018fully} and receiver~\cite{shirazi2021ultralow} implementations in $40$-nm CMOS. Both interfaces transport the same quantized tensors $(\hat{\mathbf{F}}_1,\hat{\mathbf{F}}_2)$ from Section~\ref{sec:interface}; only the physical communication layer changes.

Fig.~\ref{fig:txrx} shows both interfaces. The wireless transmitter (a) uses a
switched-capacitor DC-DC converter, an integer-N PLL, and a class-D PA; the
receiver is direct-conversion with a current-reused LNA, passive I/Q
downconversion, and analog baseband. PA supply sets output power, trading
energy for link budget. The wired interface (b) is
asymmetric~\cite{zhou2026frequencyformer}: a dynamic integrator ahead of the comparator
accumulates each bit over its period, so improved SNR lets a reduced Tx swing
meet the same BER at far lower sensor-side power. Both stages are fully
clocked, and half-period integration sustains DDR throughput.

\noindent\textbf{Duty cycling, not rate scaling.}
Both links run at a fixed rate $R_{\mathrm{PHY}}$, sending each
payload as a burst and idling between bursts:
\vspace{-1mm}
\begin{equation}
D_{\mathrm{link}} =
\frac{f B_{\mathrm{frame}}}{R_{\mathrm{PHY}}},
\qquad
fB_{\mathrm{frame}} \le R_{\mathrm{PHY}}.
\label{eq:duty}
\vspace{-1mm}
\end{equation}
Hence $s$, $d_{\mathrm{tok}}$, and $b$ set the duty cycle through
$B_{\mathrm{frame}}$, and fixed-rate spatial sampling
(Section~\ref{sec:stream}) divides it by $G$, cutting both wireless duty cycle
and wired bandwidth without changing hardware.

\iffalse

Fig.~\ref{fig:txrx} shows the wireless implementation. The transmitter uses a switched-capacitor DC-DC converter, an integer-N PLL, and a class-D PA, while the receiver employs a direct-conversion architecture with a current-reused LNA, passive I/Q downconversion, and analog baseband processing. Output power can be adjusted through the PA supply, providing an energy--link-budget tradeoff.

\noindent\textbf{Duty cycling, not rate scaling.}
The wireless link operates at a fixed physical-layer rate $R_{\mathrm{PHY}}$. Each feature payload is transmitted as a burst at this rate, with the radio remaining idle between bursts. For a payload of $B_{\mathrm{frame}}$ bits at frame rate $f$, 
\vspace{-1mm}
\begin{equation}
D_{\mathrm{link}} =
\frac{f B_{\mathrm{frame}}}{R_{\mathrm{PHY}}},
\qquad
fB_{\mathrm{frame}} \le R_{\mathrm{PHY}}.
\label{eq:duty}
\vspace{-1mm}
\end{equation}
Thus, the interface parameters $s$, $d_{\mathrm{tok}}$, and $b$ directly determine the radio duty cycle through $B_{\mathrm{frame}}$. In streaming mode, the fixed-rate spatial sampling of Section~\ref{sec:stream} further reduces the payload by a factor of $G$, proportionally reducing the wireless duty cycle and the required wired-link bandwidth without changing the underlying communication hardware.

\fi

\section{Experimental Results}
\label{sec:results}
% \vspace{-2mm}
\vspace{-1mm}
\subsection{Setup}
\vspace{-1mm}
 
\noindent\textbf{Datasets.} We evaluate static-image classification on CIFAR-10~\cite{CIFAR-10} and Visual Wake Words (VWW)~\cite{chowdhery2019visual}. For the always-on streaming case we consider Reading in the Wild (RitW)~\cite{yang2026reading} a good match for the setting CoViS targets, since the task is defined on an always-on, battery-powered, head-mounted device, which is exactly the regime in which a duty-cycled radio rather than a wired bus carries the sensor output. Following the official preprocessing, we use the gaze-centered $5^\circ$ RGB crop resized to $224{\times}224$; gaze is used only to define the crop, and head pose is excluded. At inference, the processor consumes $2$\,s windows resampled to $15$, $10$, or $5$\,FPS.
 
\noindent\textbf{Configurations.} Table~\ref{tab:main} lists the operating points used
throughout, together with everything they determine: the sensor-output rate from
\eqref{eq:rate}, the compute on each side of the link, and accuracy. They span the three
knobs of \eqref{eq:rate}: the spatial reduction $s$ that sets $N_1$, the coefficient
budget $(k_Y,k_{Cb},k_{Cr})$ that sets $d_{\mathrm{tok}}$, and the interface precision
$b$. Both stems use $L{=}8$ blocks, with the stride-2 stage folded into the depthwise
convolution of the corresponding block. The processor uses $D_s{=}96$, $A{=}2$
self-attention layers per sequence with $4$ heads and MLP ratio $4$, and the interleaved
bridge placement of Section~\ref{sec:backend}. 
 
\noindent\textbf{Training.} All CoViS configurations are trained end-to-end from scratch with quantization-aware training (QAT) with gradients passing through quantization by straight-through estimation, using AdamW~\cite{loshchilov2017decoupled} (lr $\in\{1\mathrm{e}{-3},1\mathrm{e}{-4}\}$, weight decay $1\mathrm{e}{-3}$, batch size $512$) using cosine annealing with a $10$-epoch warm-up, for $800$ epochs on CIFAR-10 and VWW and $10$ epochs on RitW.

\begin{table}[t]
\centering
\fontsize{6pt}{6.5pt}\selectfont
\setlength{\tabcolsep}{1.3pt}
\caption{CoViS operating points. All share an $8{\times}8$ block DCT, $L{=}8$ blocks per
stem, and $224{\times}224$ input. }
\label{tab:main}
\vspace{-3mm}
\begin{tabular}{l ccccc rr rr cc}
\hline
\multirow{2}{*}{cfg} & \multicolumn{5}{c}{settings} &
\multicolumn{2}{c}{transmission rate} & \multicolumn{2}{c}{MACs / frame} &
\multicolumn{2}{c}{acc. (\%)} \\
\cline{2-6}\cline{7-8}\cline{9-10}\cline{11-12}
 & $(k_Y,k_{Cb},k_{Cr})$ & $d_{\mathrm{tok}}$ & $s$ & $N_1$ & $b$
 & kb/frame  & vs.\ raw
 & sensor & processor & CF10 & VWW \\
\hline
 \multicolumn{6}{l}{\emph{reference:} raw $8$\,bpp $+$ ResNet-18}
  & $1204.2$  & $1.0\times$
  & --- & $1.82$\,G & 90.71 & 89.1 \\
\multicolumn{6}{l}{\phantom{\emph{reference:}} raw $8$\,bpp $+$ ViT-Tiny}
  & $1204.2$  & $1.0\times$
  & --- & $1.26$\,G & 90.73 & 88.6 \\
\multicolumn{6}{l}{\phantom{\emph{reference:}} raw $8$\,bpp $+$ MobileNetV3}
  & $1204.2$  & $1.0\times$
  & --- &56 M  & 89.23 & 88.2 \\
\multicolumn{6}{l}{\phantom{\emph{reference:}} FF (INT8-HAQ, 24-ch, scratch)}
  & $9.4$  & $128.0\times$
  & 26M &1.13G & 84.21 & 85.68 \\

\hline
CoVis-A & $(28,10,10)$ & 48 & 16 & 196 & 8 & $94.1$  & $12.8\times$
  & \ $10.3$\,M & $84.6$\,M & \textbf{90.11} & 89.11 \\
CoVis-B & $(28,10,10)$ & 48 & 16 & 196 & 6 & $70.6$  & $17.1\times$
  & \ $10.3$\,M  & $84.6$\,M & 90.02 & \textbf{89.14} \\
CoVis-C & $(28,10,10)$ & 48 & 32 & \ 49 & 6 & $28.2$ & $42.7\times$
  & \ $8.7$\,M  & $29.2$\,M & 89.45 & 88.61 \\
CoVis-D & $(14,5,5)$   & 24 & 32 & \ 49 & 8 & $18.8$  & $64.0\times$
  & \ $3.1$\,M & $29.0$\,M & 88.20 & 88.08 \\
\rowcolor{gray!12}
CoVis-E & $(14,5,5)$   & 24 & 32 & \ 49 & 6 & $\mathbf{14.1}$ & $\mathbf{85.3\times}$
  & \ $\mathbf{3.1}$\,\textbf{M} & $\mathbf{29.0}$\,\textbf{M}
  & 88.33 & 87.94 \\
\hline
\end{tabular}
\vspace{-5mm}
\end{table}

\begin{table}[t]
\centering
\caption{Comparison with prior in/near-sensor computing work.}
\label{tab:sota}
\vspace{-3mm}
\fontsize{6.5pt}{6.5pt}\selectfont
\setlength{\tabcolsep}{5pt}
\begin{tabular}{lccccc}
\hline
method & task & res.  & network  &TOPS/W & acc. \\
\hline
Senputing~\cite{xu2021senputing}   & MNIST& 28$^2$ & 2-layer MLP & 4.7& 93.76\\
SCAMP~\cite{bose2020fully} & MNIST& 256$^2$ &2-layer CNN & 0.535 & 93.0 \\
MR-PIPA~\cite{abedin2022mr} & MNIST & 256$^2$     &  3-layer CNN & 1.89& 97.26\\
DPCE~\cite{xu2020utilizing}       &CIFAR10& 32$^2$ & LeNet-5 &11.49  &87.20 \\
PIPSIM~\cite{roohi2023pipsim}     &CIFAR10&64$^2$ & LeNet-5 &4.12 & 90.05 \\
P$^2$M~\cite{datta2022processing} &VWW &224$^2$ & MobileNetV2 &0.4  &84.3\\
FreqFormer~\cite{zhou2026frequencyformer} & VWW &224$^2$ & ViT-Tiny (pretrained, LUT) & 28.8 & 87.80 \\
\rowcolor{gray!12}
CoViS-E & CIFAR10 & 224$^2$ &hybrid (scratch, LUT) & 24.65  & 88.33 \\
\rowcolor{gray!12}
CoViS-E & VWW & 224$^2$ &hybrid (scratch, LUT) & 24.65 & 87.94 \\
\hline
\end{tabular}
\vspace{-7mm}
\end{table}

\vspace{-2.mm}
\subsection{Accuracy}
\vspace{-1mm}
 
\subsubsection{Image Classification Tasks}

Table~\ref{tab:main} shows that CoViS approaches or exceeds the accuracy of conventional backbones while substantially reducing communication and processor cost. CoViS-A is within $0.6$ points of ViT-Tiny on CIFAR-10 and exceeds it on VWW, with $12.8\times$ fewer transmitted bits and $14.9\times$ fewer processor MACs. CoViS-B achieves the best VWW accuracy in the table ($89.14$) at $17.1\times$ compression, while CoViS-C slightly exceeds MobileNetV3 on CIFAR-10 with half the processor cost and $42.7\times$ less data. Across CoViS-A--E, payload, sensor compute, and processor compute decrease by $6.7\times$, $3.3\times$, and $2.9\times$, respectively, with only $1.78$ and $1.17$ points of accuracy loss on CIFAR-10 and VWW. INT8$\rightarrow$INT6 causes no measurable loss, whereas reducing $d_{\mathrm{tok}}$ has a larger impact, motivating $b{=}6$ for the low-rate configurations.

\subsubsection{Always-On Operation on Egocentric Video}
\label{sec}

Table~\ref{tab:stream} evaluates CoViS on RitW with fixed-rate spatial sampling ($G{=}4$). Raw video requires $18.1$, $12.0$, and $6.0$\,Mbps at $15$, $10$, and $5$\,FPS, respectively, exceeding the $1$\,Mbps wireless capacity. At $10$\,FPS, CoViS-E requires only $35.3$\,kbps and achieves $87.2\%$ accuracy, within $0.5$ points of the raw reference. Its hourly link, sensor, and processor energies are $0.65$, $0.35$, and $0.89$\,J at $-10$\,dBm, respectively. Reducing to $5$\,FPS lowers them to $0.32$, $0.18$, and $0.44$\,J/h, but reduces accuracy to $85.2\%$, favoring $10$\,FPS as the accuracy--efficiency tradeoff.

\vspace{-1mm}
\subsection{Comparison with Prior Near-Sensor Pipelines}
\vspace{-1mm}

Table~\ref{tab:sota} compares CoViS with representative near-sensor designs. Most prior work uses low-resolution workloads such as MNIST or CIFAR-10 at $32^2$--$64^2$; we focus on the two designs operating at full $224^2$ resolution. Compared with P$^2$M, CoViS-E improves VWW accuracy by $3.6\%$ while achieving $61\times$ higher energy efficiency ($24.65$ vs.\ $0.4$\,TOPS/W). Compared with Zhou et al. \cite{zhou2026frequencyformer}, CoViS-E is $0.14\%$ better on VWW at $86\%$ of its TOPS/W. Unlike Zhou et al. \cite{zhou2026frequencyformer}, which uses a pretrained ViT-Tiny and expands tokens on the host, CoViS trains the hybrid pipeline from scratch and performs a slight dimensional expansion only on the processor, avoiding the most processor-side energy cost.

\begin{table}[t]
\centering
\caption{Always-on RitW streaming with fixed-rate spatial sampling ($G{=}4$).
Energies are per hour: link/sensor/processor; $E_{\mathrm{link}}$ is reported at $0/-10$\,dBm. Rates above $1$\,Mbps are unsupported.}
\vspace{-3mm}
\label{tab:stream}
\fontsize{5.5pt}{5.8pt}\selectfont
\setlength{\tabcolsep}{2.5pt}
\begin{tabular}{llrrcccc}
\hline
FPS & cfg
& $R$ & $D_{\mathrm{link}}$
& $E_{\mathrm{link}}/\mathrm{h}$
& $E_{\mathrm{sens}}/\mathrm{h}$
& $E_{\mathrm{proc}}/\mathrm{h}$
& acc. \\
& & (kbps) & & (J) & (J) & (J) & (\%) \\
\hline

\multirow{4}{*}{15}
& raw $8$\,bpp $+$ ViT-Tiny
& $18{,}063$
& \multicolumn{2}{c}{\emph{exceeds} $R_{\mathrm{PHY}}$}
& $0.52$ & $1013.20$ & 88.2 \\
& FF (INT8-HAQ, 48-ch, scratch)
& $282.1$ & $0.282$
& $8.11/5.16$
& $0.59$ & $1008.61$ & 84.7 \\
& CoViS-C-streaming
& $105.8$ & $0.106$
& $3.04/1.93$
& $0.55$ & $1.35$ & 87.6 \\
& CoViS-E-streaming
& $52.9$ & $0.053$
& $1.52/0.97$
& $0.53$ & $1.34$ & 87.2 \\
\hline

\multirow{4}{*}{10}
& raw $8$\,bpp $+$ ViT-Tiny
& $12{,}042$
& \multicolumn{2}{c}{\emph{exceeds} $R_{\mathrm{PHY}}$}
& $0.34$ & $488.12$ & 87.7 \\
& FF (INT8-HAQ, 48-ch, scratch)
& $188.0$ & $0.188$
& $5.40/3.44$
& $0.39$ & $485.06$ & 83.6 \\
& CoViS-C-streaming
& $70.6$ & $0.071$
& $2.03/1.29$
& $0.37$ & $0.90$ & 87.4 \\
& CoViS-E-streaming
& $35.3$ & $0.035$
& $1.01/0.65$
& $0.35$ & $0.89$ & 87.2 \\
\hline

\multirow{4}{*}{5}
& raw $8$\,bpp $+$ ViT-Tiny
& $6{,}021$
& \multicolumn{2}{c}{\emph{exceeds} $R_{\mathrm{PHY}}$}
& $0.17$ & $150.39$ & 87.1 \\
& FF (INT8-HAQ, 48-ch, scratch)
& $94.0$ & $0.094$
& $2.70/1.72$
& $0.20$ & $148.86$ & 83.8 \\
& CoViS-C-streaming
& $35.3$ & $0.035$
& $1.01/0.65$
& $0.18$ & $0.45$ & 86.4 \\
& CoViS-E-streaming
& $17.6$ & $0.018$
& $0.51/0.32$
& $0.18$ & $0.44$ & 85.2 \\
\hline

\multicolumn{8}{l}{
\scriptsize All models consume a $2$-s temporal window as input.
% Raw and FF concatenate all window tokens; CoViS uses factorized space--time attention.
}
\end{tabular}
\vspace{-5mm}
\end{table}

\begin{table}[t]
\centering
\caption{Ablations at CoViS-E. Sensor MACs cover the tokenizer, stems and FiLM on CiFAR-10.
Processor default setting: $D_s{=}96$, $A{=}2$.}
\vspace{-3mm}
\label{tab:abl}
\fontsize{6.5pt}{6pt}\selectfont
\setlength{\tabcolsep}{3.4pt}
\begin{tabular}{llcccc}
\hline
\multirow{2}{*}{group} & \multirow{2}{*}{variant} & acc. & sensor & processor & trans. rate \\
 & & (\%) & MACs & MACs & (kb/frame) \\
\hline
--- & \textbf{CoViS-E (full)} & \textbf{88.33} & \textbf{3.07\,M} & \textbf{29.1\,M} & \textbf{14.1} \\
\hline
\multirow{3}{*}{multi-scale}
 & fine scale only ($\mathbf{T}_1$)   & 87.13 & $2.76$\,M & $12.2$\,M & \ $7.1$ \\
 & coarse scale only ($\mathbf{T}_2$) & 85.60 & $1.37$\,M & $12.2$\,M & \ $7.1$ \\
 & no FiLM (drop $\mathbf{t}_3$)      & 85.19 & $3.07$\,M & $29.1$\,M & $14.1$ \\
\hline
\multirow{3}{*}{fusion}
 & no cross-attn bridge          & 88.19 & $3.07$\,M & $24.4$\,M & $14.1$ \\
 & bridge before self-attn       & 88.27 & $3.07$\,M & $29.1$\,M & $14.1$ \\
 & single concatenated sequence  & 87.54 & $3.07$\,M & $25.9$\,M & $14.1$ \\
\hline
\multirow{2}{*}{stem}
 & no pre-downsample blocks   & 87.57 & $2.14$\,M & $29.1$\,M & $14.1$ \\
 & standard conv, not DWPW    & 87.97 & $14.3$\,M & $29.1$\,M & $14.1$ \\
\hline
\multirow{3}{*}{where to expand}
 & widen to $D_s$ on the sensor & \multirow{2}{*}{88.47} & \multirow{2}{*}{$8.51$\,M} & \multirow{2}{*}{$28.8$\,M} & \multirow{2}{*}{$56.4$} \\
 & \textit{(applied before the stems)} \\
 & no expansion, $D_s = d_{\mathrm{tok}}$ & 82.13 & $3.07$\,M & \ $2.3$\,M & $14.1$ \\
\hline
\end{tabular}
\vspace{-7mm}
\end{table}

\vspace{-1mm}
\subsection{Ablation Studies}
\vspace{-1mm}

Table~\ref{tab:abl} isolates the main design choices of CoViS-E on CIFAR-10. Removing the coarse or fine scale reduces accuracy by $1.20\%$ and $2.73\%$, respectively, while removing FiLM costs $3.16\%$. Eliminating or repositioning the cross-attention bridge changes accuracy by less than $0.2\%$, whereas concatenating the two streams slightly saves MACs at CoViS-E but becomes less efficient for larger $N_1$. Removing the pre-downsampling stem blocks saves $30\%$ sensor compute at a $0.76\%$ accuracy cost, while replacing depthwise-separable convolutions with standard $3{\times}3$ convolutions increases sensor compute by $4.7\times$. Finally, expanding to $D_s$ before transmission gains only $0.14\%$ accuracy while quadrupling the payload, whereas removing expansion entirely costs $6.20\%$, confirming that feature expansion is best placed on the processor side.

\begin{table}[t]
\centering

%---------------------------------------------------------- left
\begin{minipage}[t]{0.43\columnwidth}
\centering
\caption{Measured communication parameters.}
\vspace{-3mm}
\label{tab}
\fontsize{5pt}{6pt}\selectfont
\setlength{\tabcolsep}{0.8pt}
\begin{tabular}{llcc}
\hline
link & side & technology & energy/bit \\
\hline
\multicolumn{2}{l}{\textit{w/ $R_{\mathrm{PHY}}$: $1$\,Mbps}}  \\
\multirow{3}{*}{wireless}
& TX@$0$\,dBm & $40$\,nm & $4.50$\,nJ \\
& TX@$-10$\,dBm & $40$\,nm & $1.60$\,nJ \\
& RX$^{}$ & $40$\,nm & $3.50$\,nJ* \\
\hline
\multirow{2}{*}{MIPI-IR}
& TX & $65$\,nm & $0.8$\,pJ \\
& RX & $65$\,nm & $6.0$\,pJ \\
\hline
\end{tabular}
{\fontsize{4.5pt}{5pt}\selectfont $^{*}$Includes the off-chip reference-clock power.}
\end{minipage}%
\hspace{0.02\columnwidth}%
%---------------------------------------------------------- right
\begin{minipage}[t]{0.55\columnwidth}
\centering
\caption{Wireless-link cost at $10$\,FPS; $P_{\mathrm{avg}}$ is evaluated at $-10$\,dBm.}
\vspace{-3mm}
\label{tab}
\fontsize{5pt}{5pt}\selectfont
\setlength{\tabcolsep}{0.5pt}
\begin{tabular}{lcccrr}
\hline
\multirow{2}{*}{cfg}
& \multirow{2}{*}{$D_{\mathrm{link}}$}
& \multicolumn{2}{c}{$E_{\mathrm{link,frame}}$ (\textmu J)}
& $P_{\mathrm{avg}}$ & $t_{\mathrm{ser}}$ \\
& & $0$\,dBm & $-10$\,dBm & (mW) & (ms) \\
\hline
CoViS-A & $0.94$ & $752.8$ & $479.9$ & $4.80$ & $94$ \\
CoViS-B & $0.71$ & $564.8$ & $360.1$ & $3.60$ & $71$ \\
CoViS-C & $0.28$ & $225.6$ & $143.8$ & $1.44$ & $28$ \\
CoViS-D & $0.19$ & $150.4$ & $95.9$ & $0.96$ & $19$ \\
\rowcolor{gray!12}
CoViS-E & $\mathbf{0.14}$ & $\mathbf{112.8}$ & $\mathbf{71.9}$
& $\mathbf{0.72}$ & $\mathbf{14}$ \\
\hline
\multicolumn{6}{l}{\emph{with fixed-rate spatial sampling, $G{=}4$}} \\
\rowcolor{gray!12}
CoViS-E-stm. & $0.04$ & $28.2$ & $18.0$ & $0.18$ & $3.5$ \\
\hline
\end{tabular}
\end{minipage}
\vspace{-7mm}
\end{table}

\vspace{-2mm}
\subsection{Communication Link Evaluation}
\label{sec:res-link}
\vspace{-1mm}

\noindent\textbf{Energy model.}
We evaluate both wireless and wired sensor--processor communication using measured circuit-level energy. For either interface, the energy required to transfer one feature frame is
$
E_{\mathrm{link,frame}}
$=$ B_{\mathrm{frame}}
\left(E_{\mathrm{TX,bit}}+E_{\mathrm{RX,bit}}\right),
$
and the corresponding average link power is
$
P_{\mathrm{link,avg}} $=$ f E_{\mathrm{link,frame}}.
$
Thus, link energy scales directly with the transmitted representation size, while frame rate affects only average power.
For the wireless link, the measured $40$-nm TX consumes $4.5$ and $1.6$\,nJ/bit at $0$ and $-10$\,dBm, respectively. The RX consumes $3.5$\,nJ/bit, including the off-chip reference-clock power. For wired communication, we use the asymmetric MIPI integrating-receiver (MIPI-IR) interface adopted in~\cite{zhou2026frequencyformer,Int_Rx}. The $65$-nm prototype uses a reduced $\sim0.4$\,V TX swing and achieves $E_{\mathrm{TX,bit}}=0.8$\,pJ/bit and $E_{\mathrm{RX,bit}}=6.0$\,pJ/bit.

\noindent\textbf{Latency.}
For the wireless link, communication latency is dominated by payload serialization,
$
t_{\mathrm{ser}}$ =
$\frac{B_{\mathrm{frame}}}{R_{\mathrm{PHY}}}$,
where $R_{\mathrm{PHY}}$=$1$\,Mbps. Protocol-dependent framing, acknowledgement, and scheduling overheads are excluded. Wired MIPI operates at substantially higher physical-layer rates, making its serialization latency $\mu$s-scale and negligible for the workloads considered here; we therefore report latency only for the wireless link. Table~\ref{tab} summarizes the communication results. At $10$\,FPS, CoViS-E uses only a $14\%$ wireless duty cycle, reduced to $4\%$ with $G{=}4$ fixed-rate spatial sampling. At $-10$\,dBm, its link energy is $71.9$\,\textmu J/frame over wireless versus $0.096$\,\textmu J/frame with MIPI-IR, further dropping to $18.0$ and $0.024$\,\textmu J/frame in streaming mode. The same schedule reduces wireless serialization latency from $14.1$ to $3.5$\,ms.

\noindent\textbf{Technology scaling.}
Our wireless results reflect the measured $40$-nm implementation rather than the interface's energy limit. A first-order estimate suggests that halving the class-D PA switch on-resistance would increase drain efficiency from $\sim26\%$ to $\sim41\%$ and reduce TX energy from $4.5$ to $\sim3.1$\,nJ/bit at $0$\,dBm and $1$\,Mbps, assuming unchanged load and other TX overheads. For reference, Babaie \emph{et al.}~\cite{babaie2016fully} report a measured $28$-nm BLE transmitter consuming $3.6$\,mW at $0$\,dBm ($3.6$\,nJ/bit at $1$\,Mbps). Despite architectural differences, this supports the potential for further savings through technology scaling and circuit re-optimization.

\vspace{-1mm}
\subsection{System-Level Energy}
\vspace{-1mm}

We evaluate the complete sensor-to-inference energy as the sum of image readout, near-sensor computation, communication, and processor inference. Following Zhou et al.~\cite{zhou2026frequencyformer}, we use a fixed image-readout (224$\times$224$\times$3 bits at INT8) energy of $9.57$\,\textmu J/frame. For near-sensor computation, we use $45$-nm INT8 synthesis results from~\cite{sen2025look}: $154.77$\,fJ/multiplication for the conventional array-multiplier implementation (MAC) and $81.15$\,fJ/multiplication for the D\&C-based LANA implementation (LUT) \cite{sen2025lana}.
% , consistent with the evaluation methodology of FrequencyFormer. 
For wireless baselines, standard BLE uses the commercial TI CC2640R2F~\cite{ti_cc2640r2f}, whose $0$\,dBm TX and RX consume $18.3$ and $17.7$\,nJ/bit at $1$\,Mbps, respectively, for a combined $36$\,nJ/bit. The optimized $40$-nm link uses the measured TX/RX implementations of Section~\ref{sec:res-link}: including the RX off-chip reference clock, its combined energy is $7.98$\,nJ/bit at $0$\,dBm and $5.08$\,nJ/bit at $-10$\,dBm. The wired MIPI-IR interface consumes $6.8$\,pJ/bit. 
% For processor inference, we use the Jetson Orin Nano Super at $2.94$\,pJ/MAC~\cite{nvidia_orin_nano_super}.
For processor inference, we adopt a peak-performance-based energy model using the NVIDIA Jetson Orin Nano Super, which provides 17 FP16 TFLOPS under a nominal 25 W module power budget, yielding an estimated energy cost of 2.94 pJ/MAC~\cite{nvidia_orin_nano_super}.

\begin{table}[t]
\centering
\fontsize{5.8pt}{6.5pt}\selectfont
\setlength{\tabcolsep}{1.3pt}
\caption{System-level energy per frame ($\mu$J) on CIFAR-10 at $0.5$\,FPS.
Reductions are relative to the matched raw-input baseline under the same
communication interface.}
\vspace{-3mm}
\label{tab:system_energy}
\begin{tabular}{cll|cccc|rr}
\hline
& & & \multicolumn{4}{c|}{energy ($\mu$J)} & & \\
\cline{4-7}
id & cfg & link &
img.\ read. & sens. & comm. & proc. & total & red. \\
\hline

\multicolumn{9}{l}{\emph{Wired communication}} \\
M1a & raw + ViT-Tiny & MIPI-IR
& 9.57 & ---   & 8.189 & 3705.9 & 3723.6 & $1.0\times$ \\
M1b & raw + MobileNetV3 & MIPI-IR
& 9.57 & --- & 8.189 & 164.6 & 182.4 & $20.4\times$ \\
M2 & FF \cite{zhou2026frequencyformer} & MIPI-IR
& 9.57 & 1.300 & 0.064 & 3322.2 & 3333.1 & $1.12\times$ \\
M3 & CoViS-E, MAC & MIPI-IR
& 9.57 & 0.476 & 0.096 & 85.3 & 95.4 & $39.0\times$ \\
\rowcolor{gray!12}
M4 & \textbf{CoViS-E, LUT} & \textbf{MIPI-IR}
& \textbf{9.57} & \textbf{0.249} & \textbf{0.096}
& \textbf{85.3} & \textbf{95.2} & $\mathbf{39.1\times}$ \\
\hline

\multicolumn{9}{l}{\emph{Wireless communication}} \\
B1a & raw + ViT-Tiny & std.\ BLE@$0$\,dBm
& 9.57 & --- & 43352.1 & 3705.9 & 47067.5 & $1.0\times$ \\
B1b & raw + MobileNetV3 & std.\ BLE@$0$\,dBm
& 9.57 & --- & 43352.1 & 164.6 & 43526.3 & $1.1\times$ \\
B2 & CoViS-E, LUT & std.\ BLE@$0$\,dBm
& 9.57 & 0.249 & 508.0 & 85.3 & 603.1 & $78.0\times$ \\
\hline
B3 & raw + ViT-Tiny & opt.\ BLE@$0$\,dBm
& 9.57 & --- & 9609.7 & 3705.9 & 13325.2 & $3.5\times$ \\
B4 & CoViS-E, LUT & opt.\ BLE@$0$\,dBm
& 9.57 & 0.249 & 112.6 & 85.3 & 207.7 & $226.4\times$ \\
\hline
B5 & raw + ViT-Tiny & opt.\ BLE@$-10$\,dBm
& 9.57 & --- & 6117.5 & 3705.9 & 9832.9 & $4.8\times$ \\
\rowcolor{gray!12}
B6 & \textbf{CoViS-E, LUT} & \textbf{opt.\ BLE@$-10$\,dBm}
& \textbf{9.57} & \textbf{0.249} & \textbf{71.7}
& \textbf{85.3} & \textbf{166.8} & $\mathbf{281.9\times}$ \\
\hline
\end{tabular}
\vspace{-7mm}
\end{table}

Table~\ref{tab:system_energy} demonstrates the energy benefits of CoViS across wired and wireless deployments. With MIPI-IR, CoViS-E reduces processor energy to $85.3$\,\textmu J/frame, achieving $39.1\times$ and $35.0\times$ lower system energy than raw ViT-Tiny and FrequencyFormer~\cite{zhou2026frequencyformer},
respectively. Wireless deployment further amplifies these savings. Even compared with the lightweight MobileNetV3, CoViS-E reduces system energy by $72.2\times$ over standard BLE, highlighting the benefits of jointly optimizing communication and processor inference. With the optimized $-10$\,dBm link, CoViS-E achieves up to
$281.9\times$ energy reduction over raw ViT-Tiny with standard BLE, with communication and inference accounting for $43\%$ and $51\%$ of the remaining energy, respectively. 
LUT-based computation further reduces near-sensor energy from $0.476$ to $0.249$\,\textmu J/frame, primarily benefiting sensor-side power and footprint.

 \vspace{-1mm}
\section{Conclusion}
\label{sec:conclusion}
 \vspace{-1mm}

CoViS co-optimizes near-sensor feature extraction,
communication, and backend inference using compact
representations. CoViS-E reduces sensor-output bandwidth
by $85.3\times$ with only $3.1$\,M sensor-side and
$29.0$\,M processor MACs. Fixed-rate spatial sampling
further reduces wireless traffic to $35.3$\,kbps at
$10$\,FPS with minimal accuracy loss. Overall, CoViS-E
achieves up to $281.9\times$ lower system energy than
raw-frame inference over BLE and $39.1\times$
lower energy than raw MIPI-IR.

\bibliographystyle{IEEEtran}
\bibliography{Reference}\vspace{-2em}

@article{zhou2020nearsensor,
  author  = {Fei Zhou and Yang Chai},
  title   = {Near-Sensor and In-Sensor Computing},
  journal = {Nature Electronics},
  volume  = {3},
  number  = {11},
  pages   = {664--671},
  year    = {2020}
}

@article{chai2023insensor,
  author  = {Yanan Liu and others},
  title   = {In-Sensor Visual Perception and Inference},
  journal = {Intelligent Computing},
  volume  = {2},
  year    = {2023}
}

@article{xu2021senputing,
  author  = {Han Xu and others},
  title   = {Senputing: An Ultra-Low-Power Always-On Vision Perception Chip Featuring the Deep Fusion of Sensing and Computing},
  journal = {IEEE Transactions on Circuits and Systems I: Regular Papers},
  volume  = {69},
  number  = {1},
  pages   = {232--243},
  year    = {2021}
}

@inproceedings{bose2020scamp,
  author    = {Laurie Bose and others},
  title     = {Fully Embedding Fast Convolutional Networks on Pixel Processor Arrays},
  booktitle = {European Conference on Computer Vision (ECCV)},
  pages     = {488--503},
  publisher = {Springer},
  year      = {2020}
}

@article{abedin2022mrpipa,
  author  = {Minhaz Abedin and others},
  title   = {MR-PIPA: An Integrated Multilevel RRAM (HfO$_x$)-Based Processing-in-Pixel Accelerator},
  journal = {IEEE Journal on Exploratory Solid-State Computational Devices and Circuits},
  volume  = {8},
  number  = {2},
  pages   = {59--67},
  year    = {2022}
}

@inproceedings{xu2020dpce,
  author    = {Han Xu and others},
  title     = {Utilizing Direct Photocurrent Computation and 2D Kernel Scheduling to Improve In-Sensor-Processing Efficiency},
  booktitle = {Proceedings of the 57th ACM/IEEE Design Automation Conference (DAC)},
  pages     = {1--6},
  year      = {2020}
}

@article{datta2022p2m,
  author  = {Gourav Datta and others},
  title   = {A Processing-in-Pixel-in-Memory Paradigm for Resource-Constrained TinyML Applications},
  journal = {Scientific Reports},
  volume  = {12},
  number  = {1},
  pages   = {14396},
  year    = {2022}
}

@article{roohi2023pipsim,
  author  = {Arman Roohi and others},
  title   = {PIPSIM: A Behavior-Level Modeling Tool for CNN Processing-in-Pixel Accelerators},
  journal = {IEEE Transactions on Computer-Aided Design of Integrated Circuits and Systems},
  volume  = {43},
  number  = {1},
  pages   = {141--150},
  year    = {2023}
}

@article{mennel2020ultrafast,
  author  = {Lukas Mennel and others},
  title   = {Ultrafast Machine Vision with 2D Material Neural Network Image Sensors},
  journal = {Nature},
  volume  = {579},
  number  = {7797},
  pages   = {62--66},
  year    = {2020}
}

@article{wallace1992jpeg,
  author  = {Gregory K. Wallace},
  title   = {The JPEG Still Picture Compression Standard},
  journal = {IEEE Transactions on Consumer Electronics},
  volume  = {38},
  number  = {1},
  pages   = {xviii--xxxiv},
  year    = {1992}
}

@inproceedings{rao2021dynamicvit,
  title={Dynamic{V}i{T}: Efficient vision transformers with dynamic token sparsification},
  author={Rao, Yongming and others},
  booktitle={Advances in Neural Information Processing Systems},
  volume={34},
  pages={13937--13949},
  year={2021}
}

@article{bolya2023token,
  title={Token merging: Your vit but faster},
  author={Bolya, Daniel and Fu, Cheng-Yang and Dai, Xiaoliang and Zhang, Peizhao and Feichtenhofer, Christoph and Hoffman, Judy},
  journal={arXiv preprint arXiv:2210.09461},
  year={2022}
}

@article{zhou2026frequencyformer,
      title={FrequencyFormer: A Co-Designed Sensor-to-Processor Pipeline for Frequency-Domain Vision Transformer Inference}, 
      author={Chengwei Zhou and Ovishake Sen and Xuming Chen and Rishith Paramasivam and Shaahin Angizi and Swarup Bhunia and Baibhab Chatterjee and Gourav Datta},
       journal={arXiv preprint arXiv:2606.19574},
      year={2026}
}

@inproceedings{xu2020learning,
  title={Learning in the frequency domain},
  author={Xu, Kai and others},
  booktitle={2020 IEEE/CVF Conference on Computer Vision and Pattern Recognition (CVPR)},
  pages={1737--1746},
  year={2020},
  organization={IEEE}
}

@inproceedings{qin2021fcanet,
  title={Fcanet: Frequency channel attention networks},
  author={Qin, Zequn and Zhang, Pengyi and Wu, Fei and Li, Xi},
  booktitle={2021 IEEE/CVF International Conference on Computer Vision (ICCV)},
  pages={763--772},
  year={2021},
  organization={IEEE}
}

@inproceedings{gueguen2018faster,
  title={Faster neural networks straight from {JPEG}},
  author={Gueguen, Lionel and Sergeev, Alex and Kadlec, Ben and Liu, Rosanne and Yosinski, Jason},
  booktitle={Advances in Neural Information Processing Systems},
  volume={31},
  year={2018}
}

@inproceedings{ehrlich2019deep,
  title={Deep residual learning in the jpeg transform domain},
  author={Ehrlich, Max and Davis, Larry S},
  booktitle={Proceedings of the IEEE/CVF international conference on computer vision},
  pages={3484--3493},
  year={2019}
}

@inproceedings{lee2022fnet,
  title={{FN}et: Mixing tokens with {F}ourier transforms},
  author={Lee-Thorp, James and Ainslie, Joshua and Eckstein, Ilya and Ontanon, Santiago},
  booktitle={Conference of the North American Chapter of the Association for Computational Linguistics},
  pages={4296--4313},
  year={2022}
}

@article{sen2025lana,
  title={Look-up table-based energy-efficient architecture for neural accelerators ({LANA})},
  author={{Sen, Ovishake and others}},
  journal={IEEE Transactions on Computer-Aided Design of Integrated Circuits and Systems},
  volume={45},
  number={3},
  pages={1438--1452},
  year={2026}
}

@article{balle2017end,
  title={End-to-end optimized image compression},
  author={Ball{\'e}, Johannes and Laparra, Valero and Simoncelli, Eero P.},
  journal={arXiv preprint arXiv:1611.01704},
  year={2017}
}

@article{chen2016packing,
  title={Packing convolutional neural networks in the frequency domain},
  author={Wang, Yunhe and Xu, Chang and Xu, Chao and Tao, Dacheng},
  journal={IEEE transactions on pattern analysis and machine intelligence},
  volume={41},
  number={10},
  pages={2495--2510},
  year={2018},
  publisher={IEEE}
}

@inproceedings{li2023dcformer,
  title={Discrete cosin transformer: Image modeling from frequency domain},
  author={Li, Xinyu and Zhang, Yanyi and Yuan, Jianbo and Lu, Hanlin and Zhu, Yibo},
  booktitle={2023 IEEE/CVF Winter Conference on Applications of Computer Vision (WACV)},
  pages={5457--5467},
  year={2023},
  organization={IEEE}
}

@inproceedings{patro2025spectformer,
  title={{SpectFormer}: Frequency and attention is what you need in a vision transformer},
  author={Patro, Badri Narayana and others},
  booktitle={IEEE/CVF Winter Conference on Applications of Computer Vision},
  pages={3745--3755},
  year={2025}
}

@article{pan2024dctattention,
  title={Discrete cosine transform based decorrelated attention for vision transformers},
  author={Pan, Hongyi and others},
  journal={arXiv preprint arXiv:2405.13901},
  year={2024}
}

@article{lee2024dctvit,
  title={Dct-vit: High-frequency pruned vision transformer with discrete cosine transform},
  author={Lee, Jongho and Kim, Hyun},
  journal={IEEE Access},
  volume={12},
  pages={80386--80396},
  year={2024},
  publisher={IEEE}
}

@article{ibraheem2012ycbcr,
  title={Understanding color models: A review},
  author={Ibraheem, Noor A. and Hasan, Mokhtar M. and Khan, Rafiqul Z. and Mishra, Pramod K.},
  journal={ARPN Journal of Science and Technology},
  volume={2},
  number={3},
  pages={265--275},
  year={2012}
}

@article{balle2018variational,
  title={Variational image compression with a scale hyperprior},
  author={Ball{\'e}, Johannes and others},
  journal={arXiv preprint arXiv:1802.01436},
  year={2018}
}

@article{CIFAR-10,
title= {CIFAR-10 (Canadian Institute for Advanced Research)},
journal= {},
author= {Alex Krizhevsky and Vinod Nair and Geoffrey Hinton},
year= {},
url= {http://www.cs.toronto.edu/~kriz/cifar.html},
terms= {}
}

@article{chowdhery2019visual,
  title={Visual wake words dataset},
  author={Chowdhery, Aakanksha and Warden, Pete and Shlens, Jonathon and Howard, Andrew and Rhodes, Rocky},
  journal={arXiv preprint arXiv:1906.05721},
  year={2019}
}

@article{loshchilov2017decoupled,
  title={Decoupled weight decay regularization},
  author={Loshchilov, Ilya and Hutter, Frank},
  journal={arXiv preprint arXiv:1711.05101},
  year={2017}
}

@inproceedings{xu2020utilizing,
  title={Utilizing direct photocurrent computation and 2D kernel scheduling to improve in-sensor-processing efficiency},
  author={Xu, Han and others},
  booktitle={2020 57th ACM/IEEE Design Automation Conference (DAC)},
  pages={1--6},
  year={2020},
  organization={IEEE}
}

@article{datta2022processing,
  title={A processing-in-pixel-in-memory paradigm for resource-constrained tinyml applications},
  author={Datta, Gourav and others},
  journal={Scientific Reports},
  volume={12},
  number={1},
  pages={14396},
  year={2022},
  publisher={Nature Publishing Group UK London}
}

@article{sen2025look,
  title={Look-Up Table based Energy-Efficient Architecture for Neural Accelerators (LANA)},
  author={Sen, Ovishake and others},
  journal={IEEE Transactions on Computer-Aided Design of Integrated Circuits and Systems},
  year={2025},
  publisher={IEEE}
}

@ARTICLE{Int_Rx,
  author={Maity, Shovan and others},
  journal={IEEE Journal of Solid-State Circuits}, 
  title={{BodyWire: A 6.3-pJ/b 30-Mb/s -30-dB SIR-Tolerant Broadband Interference-Robust Human Body Communication Transceiver Using Time Domain Interference Rejection}}, 
  year={2019},
  volume={54},
  number={10},
  pages={2892-2906},
  doi={10.1109/JSSC.2019.2932852}}

@ARTICLE{GT_Comp,
  author={Cao, Ningyuan and others},
  journal={IEEE Journal of Solid-State Circuits}, 
  title={{A 65 nm Wireless Image SoC Supporting On-Chip DNN Optimization and Real-Time Computation-Communication Trade-Off via Actor-Critical Neuro-Controller}}, 
  year={2022},
  volume={57},
  number={8},
  pages={2545-2559},
  doi={10.1109/JSSC.2022.3159473}}

@article{photonic_in-sensor,
  title={{Integrated photonic encoder for low power and high-speed image processing}},
  author={Wang, Xiao and others},
  journal={Nature Communications},
  volume={15},
  number={1},
  pages={4510},
  year={2024},
  doi={10.1038/s41467-024-48099-2}}

@INPROCEEDINGS{ISSCC21_Comm,
  author={Okada, Chihiro and others},
  booktitle={2021 IEEE International Solid-State Circuits Conference (ISSCC)}, 
  title={{7.6 A High-Speed Back-Illuminated Stacked CMOS Image Sensor with Column-Parallel kT/C-Cancelling S\&H and Delta-Sigma ADC}}, 
  year={2021},
  volume={64},
  number={},
  pages={116-118},
  doi={10.1109/ISSCC42613.2021.9366024}}

@ARTICLE{JSSC14_Comm,
  author={Chen, Denis Guangyin and others},
  journal={IEEE Journal of Solid-State Circuits}, 
  title={{A 12 pJ/Pixel Analog-to-Information Converter Based 816 × 640 Pixel CMOS Image Sensor}}, 
  year={2014},
  volume={49},
  number={5},
  pages={1210-1222},
  doi={10.1109/JSSC.2014.2307063}}

@article{abedin2022mr,
  title={Mr-pipa: An integrated multilevel rram (hfo x)-based processing-in-pixel accelerator},
  author={Abedin, Minhaz and others},
  journal={IEEE Journal on Exploratory Solid-State Computational Devices and Circuits},
  volume={8},
  number={2},
  pages={59--67},
  year={2022},
  publisher={IEEE}
}

@inproceedings{bose2020fully,
  title={Fully embedding fast convolutional networks on pixel processor arrays},
  author={Bose, Laurie and others},
  booktitle={European Conference on Computer Vision},
  pages={488--503},
  year={2020},
  organization={Springer}
}

@inproceedings{sharifzadeh2018fully,
  title={A fully integrated multi-mode high-efficiency transmitter for IoT applications in 40nm CMOS},
  author={Sharifzadeh, Mojtaba and Masnadi-Shirazi, Amir Hossein and Rajavi, Yashar and Lavasani, Hossein Miri and Taghivand, Mazhareddin},
  booktitle={2018 IEEE Custom Integrated Circuits Conference (CICC)},
  pages={1--4},
  year={2018},
  organization={IEEE}
}

@article{shirazi2021ultralow,
  title={An ultralow-power current-reused direct-conversion Bluetooth-low-energy receiver front-end in 40-nm CMOS},
  author={Shirazi, Amir Hossein Masnadi and Lavasani, Hossein Miri and Sharifzadeh, Mojtaba and Rajavi, Yashar and Mirabbasi, Shahriar and Taghivand, Mazhareddin},
  journal={IEEE Transactions on Microwave Theory and Techniques},
  volume={69},
  number={5},
  pages={2697--2711},
  year={2021},
  publisher={IEEE}
}

@article{yang2026reading,
  title={Reading recognition in the wild},
  author={Yang, Charig and others},
  journal={Advances in Neural Information Processing Systems},
  volume={38},
  pages={141466--141485},
  year={2026}
}

@article{babaie2016fully,
  title={A fully integrated Bluetooth low-energy transmitter in 28 nm CMOS with 36\% system efficiency at 3 dBm},
  author={Babaie, Masoud and others},
  journal={IEEE Journal of Solid-State Circuits},
  volume={51},
  number={7},
  pages={1547--1565},
  year={2016},
  publisher={IEEE}
}

@misc{nvidia_orin_nano_super,
  author       = {Suhas Hariharapura Sheshadri and Leela Subramaniam Karumbunathan and Dustin Franklin},
  title        = {NVIDIA Jetson Orin Nano Developer Kit Gets a ``Super'' Boost},
  howpublished = {NVIDIA Technical Blog},
  year         = {2024},
  month        = dec,
  note         = {Accessed: 2026-09-19},
  url          = {https://developer.nvidia.com/blog/?p=93942}
}

@manual{ti_cc2640r2f,
  title        = {{CC2640R2F SimpleLink Bluetooth 5.1 Low Energy Wireless MCU Datasheet}},
  organization = {Texas Instruments},
  number       = {SWRS204C},
  year         = {2020},
  month        = sep,
  note         = {Rev. C},
  url          = {https://www.ti.com/lit/ds/symlink/cc2640r2f.pdf}
}
\end{document}